# A Layered Modeling and Simulation Approach to investigate Resource-aware Computing in MPSoCs

Aurang Zaib, Prashanth Raju, Thomas Wild, Andreas Herkersdorf
Technical University Munich, Germany
{aurang.zaib, prashanth.raju, thomas.wild, herkersdorf}@tum.de

*Abstract*—Increasing complexity of modern multi-processor system on chip (MPSoC) and the decreasing feature size have introduced new challenges. System designers have to consider now aspects which were not part of the design process in past times. Resource-aware Computing is one of such emerging design concerns which can help to improve performance, dependability and resource utilization of overall system. Resource-aware execution takes into account the resource status when executing tasks on MPSoCs. Exploration of resource-aware computing at early design stages of complex systems is mandatory and appropriate methodologies to do this in an efficient manner are thus required. In this paper, we present a modular approach which provides modeling and simulation support for investigation of resource-aware execution in MPSoCs. The proposed methodology enables rapid exploration of the design space by modeling and simulating the resource-awareness in a separate layer while widely reusing the legacy system model in the other layer. Our experiments illustrate the benefits of our approach for the exploration of resource-aware execution on MPSoCs.

## I. INTRODUCTION

Increasing transistor densities allow more and more transistors to be integrated on a single chip. The higher integration densities have led to the evolution of modern multiprocessor system on chip (MPSoC). MPSoCs are nowadays integral part of advanced embedded systems, such as powerful gaming consoles, high resolution displays and smart phones. In past times, designers of these systems were typically faced with design requirements regarding performance, flexibility, cost and power consumption.

This paradigm of multiprocessor-based compute architectures has also revolutionized the thinking of embedded system designers. In addition to the classical requirements, aspects which were not considered to be the part of a conventional design cycle have gained importance for researchers. Resource-aware computing is one of those potential aspects [1]. It performs task mapping decisions while taking into account the application demands and status of underlying system. In addition, the applications have the ability to adapt their computation according to the available resources. This allows concurrent applications to efficiently utilize the underlying platform at system run-time.

In order to investigate the design concerns like resource-aware computing, system designers require methodologies that enable early exploration of systems. Conventional monolithic simulation models require significant modeling and simulation effort while investigating the design of the complex systems with such concerns. To reduce this effort, we propose an approach that allows maximum reuse of conventional system models and simulators to evaluate the new design aspect. To reuse an existing simulator for investigating an emerging concern like resource-aware execution, a simulation concept is helpful which enables to keep the new aspect of the system functionality separated while exploring the design space. This resembles the "separation of concerns" principle which is a crucial component of modern system-level design paradigm [2]. It enables the rapid exploration of the system design space by minimizing the modeling effort and increasing the simulation speed at initial design stages.

In resource-aware computing, the decisions to utilize the platform resources do not bring any change in the functional execution which follows these decisions. For example, the resource-aware decisions to use platform resources in an efficient manner do not affect the actual functionality to be executed on underlying resources. Therefore, the non-functional execution related to the resource-awareness can be separated from the actual computation on the architecture. We have followed the separation of concerns approach which exploits this separation to enable rapid exploration of the system design space. In this paper, we propose a simulation concept, which provides modeling and simulation support for early exploration of different aspects of resource-aware computing in MPSoCs. In our experiments, we have evaluated the impact of different resource assignment strategies on architectural resources to depict the usefulness of our concept.

The remainder of the paper is organized as follows: The next section discusses related work. The details of resource-aware execution are given in the Section III. Section IV describes modeling of resource-aware execution with the proposed approach. Section V provides further details of the simulation concept. Section VI explains the simulation set up for our experiments. In Section VII, the evaluation of resource-aware execution with our simulation approach is presented. Finally, Section VIII concludes the paper.

## II. RELATED WORK

In past times, designers have used the separation of concerns principle to separate different aspects of MPSoC design such as application and architecture, or computation and communication.

The Metropolis framework [3] was amongst the first simulation environments leveraging the separation of concerns principle to address the complexity of system level design. This approach introduced the concept of meta model semantics, which represent different aspects of a design at a desired abstraction. The meta models from one aspect can be reused to allow the design space exploration of other aspects in a unified way. This work illustrates that separating different concerns offers advantages like reduction in design time and easier exploration of design space as compared to a monolithic design approach.

To investigate many aspects of embedded system design in a unified simulation framework, researchers have proposed a model-based integrated simulation (MILAN) framework [4]. In this concept, each aspect of the design, for example application, resource or communication is represented by a different model. The simulation framework facilitates rapid performance evaluation of embedded systems at different levels of abstraction. That model-based approach is applied by integrating two different tools operating at different abstraction levels, i.e. DESERT and HiPerE in the MILAN framework [5]. The above-mentioned tools work in a coordinated manner to explore a large design space.

There are state-of-the art commercial and academic simulation frameworks, which enable integration of custom models for design space exploration of a given design concern in MPSoCs [6],[7]. In order to investigate the complete system containing models of resource-aware computing and other platform specific execution by deploying separation of concerns principle, it is essential that the control flow between the resource-aware decision making and the rest of the execution is captured and simulated by keeping the simulation overhead in a reasonable bound. Custom models related to the resource-aware computing have unique interface requirements. Therefore, the interaction between resource-aware and other platform-specific models can introduce large simulation overhead, when integrated in the above-mentioned simulation frameworks. In our approach, we separate the modeling of resource-aware execution including the associated control flow and the rest of the application representing the actual functionality, in two different layers. During simulation, the layers take care for the execution of their respective parts of the model, while the control flow between the layers is supported by our simulation framework without degrading the simulation performance by great deal.

## III. RESOURCE-AWARE COMPUTING

Resource-aware execution represents a computing paradigm, in which a given application executes on the underlying platform, keeping in view the status of the platform resources. In future many-core era, resource-aware execution is under investigation by researchers, who evaluate efficient methodologies for platform utilization. In our work, we have modeled the resource-aware execution following the paradigm named as *Invasive Computing* [1]. In Invasive Computing, the resource-aware applications compete to explicitly acquire the platform resources, which are suitable for their execution. Applications specify their computation demands in terms of underlying platform resources. A resource management layer evaluates these demands from concurrent applications and allocates the resources according to certain resource management policies. The resource management policies have a direct impact on the platform utilization and execution time when different applications contend for the resources on the underlying architecture [8]. The applications can adapt themselves according to the number and type of resources which are made available to them at run-time. When the execution is finished, the applications release the acquired resources.

In Invasive Computing, *invade, assort, infect and retreat* commands are used to describe the resource-aware execution. We have represented the above-mentioned commands in the form of following execution phases in this paper.

*1) get_resource:* In the $get\_resource$ phase of resource-aware execution, the application inquires which resources in the underlying architecture comply with its computation demands. It is represented by the following equation:

$$R := get\_resource(D) \qquad (1)$$

$D$ represents set of application demands/constraints in terms of resource status. The application requirements may consist of one or more parameters related to resource status in underlying architecture, for example the CPU load, CPU temperature or reliability profile of CPU etc. $R$ represents the set of resource(s) in the underlying platform, which satisfy the application demands D.

*2) reserve_resource:* $reserve\_resource$ represents the second phase of the resource-aware execution. reserve_resource performs further sorting, to select the most appropriate resource(s) from the list of suitable resources R. The reason for performing this second level of selection after get_resource, is to sort out only that number of resources, which are requested by the application. Later the selected resources are reserved by a reservation function $res()$ and returned back to the application as the claim $C$. In case of successful reservation, claim C is given by the following equation:

$$C := \{c_i | c_i = res(r_i), r_i \in R \land r_i \text{ is successfully reserved}\} \qquad (2)$$

$c_i$ represents the resource IDs in the underlying architecture, which are successfully reserved for the demanding application. $r_i$ represents an arbitrary resource, which satisfies application demands. If there is no resource which meets the application demands, the claim is returned empty. The reserve_resource phase, in particular the reservation function res() is specific to the underlying architecture and the application constraints. In the current case-study, the scenarios are considered, in which the applications demand exclusive reservation of resources for themselves.

*3) execute_resource:* Depending on the outcome of the reserve_resource phase, $execute\_resource$ is started on the underlying platform resources. An application can carry out an arbitrary longer functional execution on the resources, which are reserved until explicitly released through $release\_resource$ phase.

*4) release_resource:* The semantics of release_resource are essentially the opposite of the actions in reserve_resource. In this phase, the acquired resources in the claim are returned to the same state as they were before the reservation.

## IV. PROPOSED METHODOLOGY

Keeping in view the separation between the resource-aware execution and the following functional execution in resource-aware computing as detailed in the Section I, we have proposed the modeling and simulation concept in which the overall execution is divided in two sub-categories: 1) *Resource-aware execution* and 2) *Functional execution*. Resource-aware execution describes the decision making to utilize the architecture resources according to the application demands keeping in view the status of underlying resources. Functional execution represents the execution other than the resource-aware execution on the selected platform resources. The decisions made during resource-aware execution do not affect the actual functional execution.

To simulate the complete system, in which the resource-aware processing and the functional execution are modeled separately, the framework has to take care of two important aspects: 1) the control flow between the two separations must be captured and modeled appropriately and 2) the captured control flow is supported during simulation execution. Our approach addresses both of these aspects. The proposed framework is shown in the Figure 1.

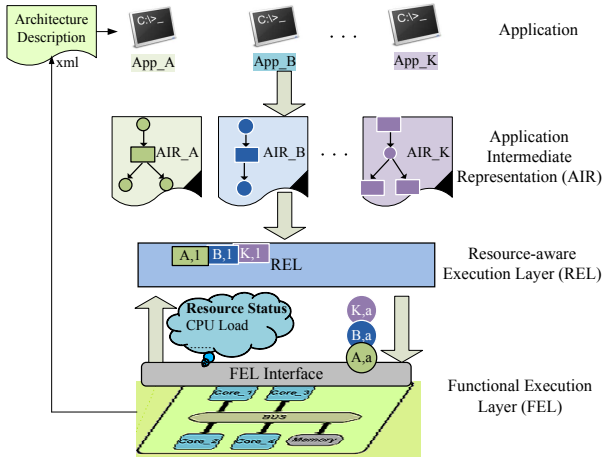

Fig. 1. Proposed framework to simulate resource-awareness

The components of the proposed framework are described in the following text.

*A. Application Intermediate Representation*

*Application Intermediate Representation (AIR)* captures the behavior of an application in a way that the resource-aware execution is separated from the functional execution. Each resource-aware application (App_$i$) is represented by an AIR (AIR_$i$). AIRs are generated in a semi-automatic way with the help of a tool, which analyzes resource-aware behavior inside invasive applications and represents it in the form of control flow graphs. A control flow graph (CFG) depicting a fragment of an AIR as shown in Figure 2.

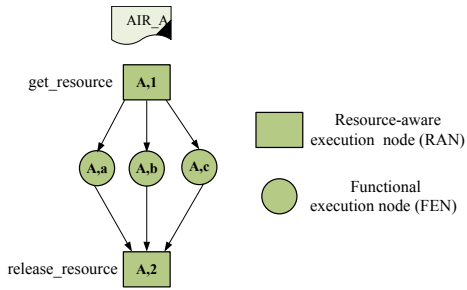

Fig. 2. Fragment of an AIR representing resource-aware application

An AIR consists of two types of nodes:
- Resource-aware execution node (RAN): This node carries the resource-aware behavior of the applications as described in the Section III. RAN nodes can be classified in further two types:
  - get_resource: This is the resource-aware node, which requests to acquire underlying resources by specifying application demands.
  - release_resource: This node releases the resources which were acquired before by the application through get_resource node.
- Functional execution node (FEN): This node consists of resource-specific execution, which is to be executed on the functional execution layer. For example, we have used application traces which can be executed on RISC processors in MPSoC system as FEN. The details of these application traces are provided later in the Section VI.

The edges of an AIR represent the resource-aware control flow. These edges capture that under which conditions which functional execution nodes shall be followed by resource-aware execution nodes. Such a representation of applications limits the modeling effort by enabling reuse of existing application models, which can be simulated on the platform to investigate resource-awareness.

*B. Resource-aware Execution Layer*

The *resource-aware execution layer (REL)* models the functionality to simulate the resource-aware processing. The details of the models which are built in this layer are given in the Section V. Resource-aware execution affects the decisions describing which parts of the application are to be executed on which underlying resources according to the application demands. These decisions are dependent on the status of resources in the functional execution layer.

*C. Functional Execution Layer*

The *functional execution layer (FEL)* is the layer on which the functional execution is carried out. We have used an MPSoC architectural simulator as FEL [9] as detailed in Section VI.

*D. FEL Interface*

The *FEL Interface* represents the extensions which we have introduced in the functional execution layer to enable its interface with REL. We have introduced monitor models which gather resource utilization data from functional execution layer and then provided it to REL for resource-aware processing.

## V. INTERPLAY OF RESOURCE-AWARE AND FUNCTIONAL EXECUTION LAYERS

In this section, we have provided the details that how the resource-execution layer is precisely modeled and how it interacts with functional execution layer to simulate the overall system. This section is divided in two sub-sections which describe the details of the components implemented inside the resource-execution layer and FEL interface respectively.

## A. Resource-aware Execution Layer

*Resource-aware Executor* and *Resource Manager* are the modules which are modeled in the REL to simulate the resource-aware execution and establish its interface with functional execution layer.

*1) Resource-aware Executor:* The resource-aware executor is responsible for interpreting AIR at simulation run-time. In order to support simulation of concurrent resource-aware applications, resource-aware executor is dynamically started per application instance to interpret the associated AIR, i.e. to take care for the dispatching of resource-aware and functional execution nodes. The implementation of resource-aware executor is independent of the application type, i.e. it can generically parse any given AIR. Depending on the type of node in AIR, resource-aware executor interacts with either the resource manager or the FEL interface accordingly. For resource-aware execution nodes, it analyses the resource-aware requirements contained inside them. If a get_resource node is encountered, resource-aware executor interacts with the resource manager which in turn gets the resource information of the underlying architecture. Based on this information, resource manager returns the claim which carries the resource IDs, made available to the application.

Depending on the outcome of RAN, the functional execution nodes are mapped on the acquired hardware resources. Resource-aware executor provides the necessary information about the FENs (FEN IDs and associated AIR IDs) to the FEL interface, to keep track of their execution on underlying resources. When an FEN node finishes its execution, the FEL interface notifies this event to the associated resource-aware executor. This is required to follow the control flow of the AIR. If release_resource node is encountered in AIR, resource-aware executor asks the resource manager to release the resources.

*2) Resource Manager:* The resource manger (RM) is a centralized instance modeled in the resource-aware execution layer to make resource allocation decisions by keeping in view the status information of the underlying architecture. In our current investigations, we have considered the status of the processing resources in the underlying architecture as a metric for resource allocation. In response to the get_resource node, RM selects the suitable resources among the list of available resources, according to the application demands. In addition, RM reserves these resources exclusively for the demanding application. In case of successful reservation, the claim is returned to resource-aware executor. When resource-aware executor calls RM for a release_resource node, the resources reserved during the execution of get_resource node are released.

## B. FEL Interface

Two components, namely the *Execution Controller* and the *Status Collector* are modeled in the FEL interface to establish the communication between resource-aware and functional execution layers.

*1) Execution Controller:* The execution controller receives the functional execution nodes in AIR and forwards them for their execution to the respective resources in FEL, i.e. CPU cores. For each arriving node, a unique context is created which allows to distinguish concurrent nodes from different applications. Each node is registered with the status *execution_started* till the time it finishes its execution. Eventually, when the execution of a node is finished, its status is changed to *execution_finished* and this information is notified to the corresponding application through resource-aware executor. In this way, the execution controller helps to simulate the control flow required between the resource-aware and functional execution layers during simulation run-time.

*2) Status Collector:* This component is responsible to acquire the desired resource information from functional execution layer and provide it to REL for resource-aware processing. Status collector collects the resource information from the monitor models built inside FEL. When RM inquires about the status of resources, *send_resource()* is called and the status information is provided to the resource manager. The interplay between the resource-aware execution layer and the FEL interface showing different resource-aware execution phases is shown in the Figure 3.

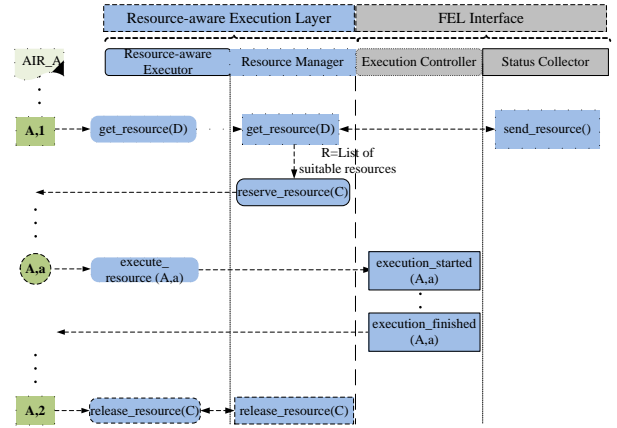

Fig. 3. Interplay between resource-aware execution layer and FEL interface

## VI. EXPERIMENT

Resource management strategies in resource-aware computing play an important role for defining the performance of overall system as presented in [8]. Therefore, we have selected two such strategies, which will be detailed later in this section, to be explored through our simulation framework. In principle, there could be many resource management schemes which highlight different aspects of platform exploration. However, the presented scenarios depict the ability of the proposed approach to enable rapid investigation of different resource management policies.

We have adapted the architectural simulator [9] as functional execution layer to investigate resource-aware computing. The used simulator is SystemC-based simulation tool for system-level performance evaluation of MPSoC architectures. In this tool, the applications are modeled in the form of so called application traces. The trace-based approach captures the internal functionality of the architecture at higher abstraction level [10]. To obtain such traces, the functional parts of the application are executed on a cycle accurate simulator of the target CPU. Describing further details about the trace-based simulation approach is out of scope of this paper.

The architecture description containing the configuration of the underlying hardware architecture is specified using xml file. For example, the designer can set the number of CPUs, parameters of the on-chip interconnect and the memory hierarchy. In the current investigations, the simulated hardware platform consists of 6 Leon3 RISC CPUs of sparcV8 architecture [11], each operating at 100 MHz and are connected via a shared bus. Each CPU has a private L1 cache of size 32kb with line size of 1024b.

For our experiments, we have used two concurrent resource-aware applications ; audio equalization and harris corner detection. Audio equalization performs equalization of 1024 samples of 6 channel audio at 44KHz and the harris corner detection processes 5 video frames each of resolution 320x240 pixels. Both applications execute on their input workload periodically with a configurable time period. In current investigation, the iteration period of audio equalization and corner detection is set to 400ms and 1000ms respectively. Before beginning their execution in each iteration, they ask for the processing resources from the resource manager which returns the resources according to available resources and a certain resource management policy. Each application has the ability to adapt its execution according to the number of CPUs allocated by the RM. For instance, when sufficient number of CPUs are not available, the audio equalization processes at reduced sample rate and the corner detection adapts its algorithm to reduce processing in order to meet the execution time deadline.

The applications are transformed into their respective AIRs. The resource-aware execution nodes in the AIRs capture the demands of the applications to acquire or release the processing resources in the underlying platform. Each RAN can have multiple control flows depending on how many processing cores were returned by RM. Whereas the functional execution nodes are the application traces which perform either audio equalization on voice samples or the corner detection on an input image frame depending upon the respective application. The traces for FENs were obtained by executing the functional parts of application code on a cycle-accurate model of Leon3 processor.

In our experiments, an application can acquire at maximum 5 CPUs for its execution. This is done to prevent starvation of any application which is not able to acquire resources because they are already reserved exclusively for other application(s) being executed over the architecture. Once the application gets the resources, it starts its execution over them. In the current investigations, we have assumed that the application follows run-to-completion execution model, i.e. once the execution is started on the allocated resources, it can not be interrupted. Therefore, it is assumed that the RM has prior information about the periodicity of the applications which is taken into account to perform resource allocation in a consistent manner. The resource management scenarios which we have considered are given in the following text.

### A. Scalability-based resource management

In first approach, RM evaluates the scalability graphs of the applications to decide the resource allocation between the two applications, i.e. higher number of cores are assigned to the application which has better scalability. Scalability graph presents the achieved execution speed up of an application for a given CPU count normalized by the execution speed on a single CPU. The scalability information of each application is provided by the programmer to the RM at the simulation startup. The scalability graphs of the two considered applications are shown in the Figure 4.

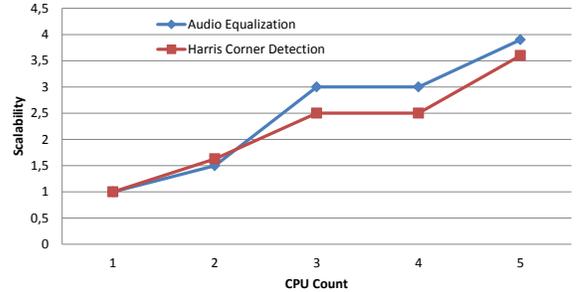

Fig. 4. Scalability graphs of the two applications

In the scalability graph, x-axis represents the number of CPUs allocated for the application. The y-axis indicates the scalability achieved when the application is alone executed on the platform i.e. there is no other contending application.

### B. Load-based resource management

In the second approach, each application provides information about the load on CPUs when mapped alone on the architecture. RM then performs resource allocation based on the standalone load information to ensure better platform utilization, i.e. the application which puts higher load on the CPUs gets higher number of resources. The standalone CPU load generated by each application is shown in the Figure 5.

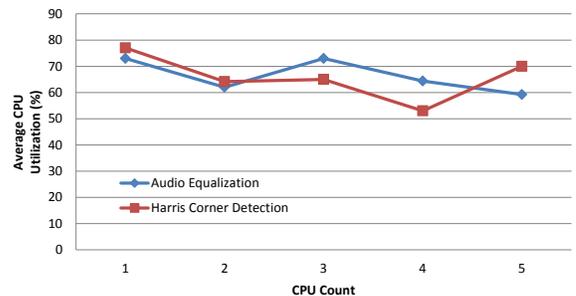

Fig. 5. Standalone CPU load of the two applications

The x-axis represents the number of CPUs allocated for the application and the y-axis indicates the average CPU utilization by the application in a standalone environment. In the next section, the simulation results for the two investigated scenarios are described.

## VII. RESULTS

In the presented results, the simulated time for each scenario was 3500 ms. In addition, each scenario is simulated 5 times and the results have been averaged over all 5 simulation runs. The CPU load on each CPU for both investigated scenarios is shown in the Figure 6.

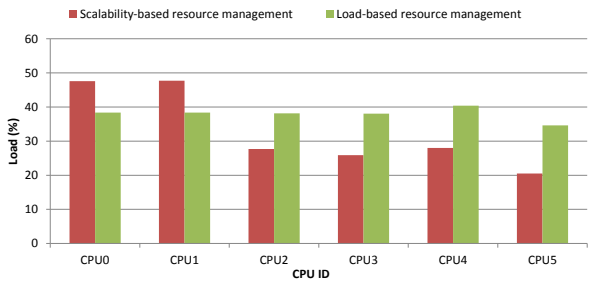

Fig. 6. CPU load for the two resource management scenarios

The average CPU load for the scalability-based and load-based scenarios is 32.89% and 37.98% respectively. Load-based scenario leads to higher average CPU utilization because of the different CPU allocation by the RM to the two application, when both of them are requesting for resources at the same time. In scalability-based approach, each application gets 3 CPUs as the scalability (speed-up) achieved by each of them is nearly the same for this CPU count as shown by the Figure 4. Whereas in load-based scenario, the RM assigns 5 cores to the more demanding application (corner detection), as influenced by the standalone load shown in Figure 5. In this case, audio equalization has to confine its execution to 1 CPU. Hence, higher number of CPU allocation to heavier load corner detection application leads to better platform utilization in case of load-based scenario. However, it is important to note that better platform utilization is achieved at the cost of performance degradation of audio equalization application. Performance degradation could either be in the form of processing at lower sampling rate or higher execution time. Number of cache accesses for each CPU for both considered scenarios are shown in the Figure 7.

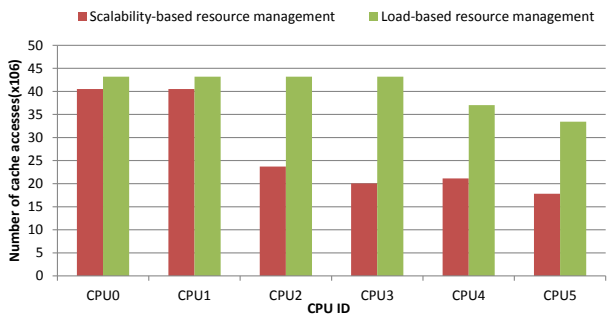

Fig. 7. Number of cache accesses for the two resource management scenarios

Again, we see that the load-based scenario stresses the system more which can be observed by comparing the number of cache accesses made in both scenarios. This goes in-line with our discussion for CPU load results. Our simulation framework enables logging of many architectural parameters like bus load, individual task execution time and the scheduler activities which allows meaningful insight of the architecture while exploring different design points.

Changing from scalability-based scenario to load-based scenario requires to alter few lines of code in the resource manager and xml configuration file. Hence, the models related to the resource-aware execution can be quickly modified without changing other system models which clearly depicts the usefulness of our approach as compared to conventional monolithic simulators. In addition, the average simulation time for scalability and the load-based scenarios for 5 simulation runs are 37.66s and 42s respectively. Therefore, different scenarios related to resource-awareness can be quickly explored in our simulation framework in very less time.

## VIII. Conclusion and Outlook

We have proposed an approach which allows rapid exploration of resource-aware computing in MPSoC architectures. Our approach makes use of separation of concerns principle to separate the decision making related to the resource-awareness from other platform-specific execution. The resource-aware execution is supported by appropriate models as proposed in our simulation concept. We have demonstrated the usefulness of our approach by investigating different resource management scenarios of resource-aware computing. The results show that our approach performs efficient exploration of the design space while simulating concurrent resource-aware applications for different resource management policies. In addition, our simulation concept helps to quickly identify candidate design points by significantly reducing the modeling and simulation effort for resource-aware computing.

## Acknowledgment

This work was partly supported by the German Research Foundation (DFG) as part of the Transregional Collaborative Research Centre Invasive Computing (SFB/TRR 89).

## References

[1] J. Teich, J. Henkel, A. Herkersdorf, D. Schmitt-Landsiedel, W. Schröder-Preikschat, and G. Snelting, "Invasive computing: An overview," in *Multiprocessor System-on-Chip*, pp. 241–268, Springer, 2011.
[2] K. Keutzer, A. Newton, J. Rabaey, and A. Sangiovanni-Vincentelli, "System-level design: orthogonalization of concerns and platform-based design," in *IEEE Transactions on Computer-Aided Design of Integrated Circuits and Systems*, pp. 1523–1543, December 2000.
[3] F. Balarin, Y. Watanabe, H. Hsieh, L. Lavagno, C. Passerone, and A. Sangiovanni-Vincentelli, "Metropolis: An integrated electronic system design environment," *Computer*, vol. 36, no. 4, pp. 45–52, 2003.
[4] A. Bakshi, V. K. Prasanna, and A. Ledeczi, "Milan: A model based integrated simulation framework for design of embedded systems," in *Proceedings of the 2001 ACM SIGPLAN workshop on Optimization of middleware and distributed systems*, August 2001.
[5] S. Mohanty, V. K. Prasanna, S. Neema, and J. Davis, "Rapid design space exploration of heterogeneous embedded systems using symbolic search and multi-granular simulation," in *Proceedings of the joint conference on Languages, compilers and tools for embedded systems: software and compilers for embedded systems*, July 2002.
[6] A. D. Pimentel, C. Erbas, and S. Polstra, "A systematic approach to exploring embedded system architectures at multiple abstraction levels," *Computers, IEEE Transactions on*, vol. 55, no. 2, pp. 99–112, 2006.
[7] Intel Corporation,, "CoFluent Studio." https://www.cofluentdesign.com/.
[8] S. Kobbe, L. Bauer, D. Lohmann, W. Schröder-Preikschat, and J. Henkel, "Distrm: distributed resource management for on-chip many-core systems," in *Proceedings of the seventh IEEE/ACM/IFIP international conference on Hardware/software codesign and system synthesis*, pp. 119–128, ACM, 2011.
[9] R. Plyaskin and A. Herkersdorf, "A method for accurate high-level performance evaluation of mpsoc architectures using fine-grained generated traces," in *Architecture of Computing Systems-ARCS 2010*, pp. 199–210, Springer, 2010.
[10] T. Wild, A. Herkersdorf, and G.-Y. Lee, "Tapestrace-based architecture performance evaluation with systemc," *Design Automation for Embedded Systems*, vol. 10, no. 2-3, pp. 157–179, 2005.
[11] Aeroflex Gaisler, "LEON 3." http://www.gaisler.com/leonmain.html.